\documentclass[12pt]{article}
\usepackage{scicite}
\usepackage{times}
\usepackage{xcolor}
\usepackage[version=4]{mhchem}
\usepackage{amsmath,stmaryrd,graphicx,amssymb}
\usepackage{mathtools}  
\usepackage{xfrac}  
\usepackage{nameref}
\usepackage{miller}
\usepackage{xr}
\usepackage[hidelinks]{hyperref}
\usepackage{gensymb}
\emergencystretch=\maxdimen
\usepackage[labelfont=bf]{caption}

\makeatletter
\renewcommand{\@seccntformat}[1]{%
  \ifcsname prefix@#1\endcsname
    \csname prefix@#1\endcsname
  \else
    \csname the#1\endcsname\quad
  \fi}

\makeatother

\usepackage[a4paper,
            left=0.75in,
            right=0.75in,
            top=0.75in,
            bottom=1in,
            footskip=.25in]{geometry}
            
\newenvironment{sciabstract}{%
 \setlength{\parindent}{0pt}\bf}

\title{Bridging experiment and theory of relaxor ferroelectrics at the atomic scale with multislice electron ptychography}

\author{Menglin Zhu$^{1\ast}$, Michael Xu$^{1\ast}$, Yubo Qi$^{2,3}$, Colin Gilgenbach$^{1}$, Jieun Kim$^{4,5}$, \\ Jiahao Zhang$^{3}$, Bridget R. Denzer$^{1}$, Lane W. Martin$^{4,6}$, Andrew M. Rappe$^{3}$,\\ James M. LeBeau$^{1\ast\ast}$\\\\
\normalsize{$^{1}$Department of Materials Science and Engineering, Massachusetts Institute of Technology,}\\
\normalsize{Cambridge, MA 02139, USA}\\
\normalsize{$^{2}$Department of Physics, University of Alabama at Birmingham, Birmingham, AL 35294, USA}\\
\normalsize{$^{3}$Department of Chemistry, University of Pennsylvania, Philadelphia, PA 19104, USA}\\
\normalsize{$^{4}$Department of Materials Science and Engineering,}\\
\normalsize{University of California, Berkeley, CA, 94720, USA}\\
\normalsize{$^{5}$Department of Materials Science and Engineering,}\\
\normalsize{Korea Advanced Institute of Science and Technology, Daejeon, 34141, Republic of Korea}\\
\normalsize{$^{6}$Departments of Materials Science and NanoEngineering, Chemistry, and Physics and Astronomy}\\
\normalsize{and the Rice Advanced Materials Institute, Rice University, Houston, TX 77005, USA}\\
\\
\normalsize{$^\ast$These authors contributed equally to this work.}\\
\normalsize{$^{\ast\ast}$To whom correspondence should be addressed; E-mail: lebeau@mit.edu.}}

\date{}

\begin{document}

\maketitle
\newpage
\begin{sciabstract}

Introducing structural and/or chemical heterogeneity into otherwise ordered crystals can dramatically alter material properties. Lead-based relaxor ferroelectrics are a prototypical example, with decades of investigation having connected chemical and structural heterogeneity to their unique properties. While theory has pointed to the formation of an ensemble of ``slush''-like polar domains, the lack of direct, spatially resolved volumetric data comparable to simulations presents a significant challenge in measuring the spatial distribution and correlation of local chemistry and structure with the physics underlying relaxor behavior. Here, we address this challenge through three-dimensional volumetric characterization of the prototypical relaxor ferroelectric \ce{0.68Pb(Mg$_{1/3}$Nb$_{2/3}$)O3-0.32PbTiO$_3$} using multislice electron ptychography. Direct comparison with molecular dynamics simulations reveals the intimate relationship between the polar structure and unit-cell level charge imbalance induced by chemical disorder. Further, polar nanodomains are maintained through local correlations arising from residual short-range chemical order. Acting in concert with the chemical heterogeneities, it is also shown that compressive strain enhances out-of-plane correlations and ferroelectric-like order without affecting the in-plane relaxor-like structure. Broadly, these findings provide a pathway to enable detailed atomic scale understanding for hierarchical control of polar domains in relaxor ferroelectric materials and devices, and also present significant opportunities to tackle other heterogeneous systems using complementary theoretical and experimental studies.
\end{sciabstract}

\newpage
\section*{Main Text}

Disruption of long-range order through heterogeneity ``can drastically modify the behaviours predicted from [models with a perfect crystalline lattice] ... but measuring this local order [and disorder] and correlating it with the microscopic physics underlying material behaviour remains a grand challenge'' \cite{Krogstad2018-gl}. This concept, used to describe the study of lead-based relaxor ferroelectrics, encapsulates decades of investigations relating results from structural characterization to unique macroscopic properties - such as high electromechanical coupling and broad, frequency- and temperature-dependent dielectric response \cite{Cohen2006-ue,Cowley2011-is}. While the consensus is that local chemical and structural inhomogeneities promote the relaxor response, several differing interpretations have been proposed based on numerous experimental and theoretical results, including the early model of polar nanoregions \cite{Burns1983-sy, Cross1987-wy} to the more recent polar nanodomain (PND) slush \cite{Takenaka2017-sj}.  

Depending on the methods and assumptions applied, experiments and theory often paint only a partial view of the whole picture. As Takenaka et al. describe, this is reminiscent of the ``legend about seven blind men and an elephant,'' in which different methods provide various, though not necessarily contradictory, physical insights into relaxors \cite{Takenaka2018-px}. This difference is seen in comparing scattering experiments and theoretical simulations, in which the results from characterization cover the averaged distribution of heterogeneities within the crystal, while simulations provide a direct atomic scale structure. Despite aligning in certain aspects, differences in the resolution and scale of structural information preclude direct comparison of nanoscale polar correlations \cite{Takenaka2017-sj, Eremenko2019-wn}, chemical inhomogeneities \cite{Hilton1990-eu, Burton2005-mf, Bokov2011-jv}, or structural distortions \cite{Chen2000-tg, Kumar2021-gu}, all of which can influence local structure and macroscopic material response. Even further, determining the structure of relaxors in the context of such heterogeneities becomes increasingly complex when considering the influence of external stimuli like temperature, boundary conditions, or electric field.

To address the ``elephant'' conundrum of relaxor materials, experimental studies providing spatially resolved volumetric structure and chemistry, comparable to the scale and information offered by molecular dynamics (MD) simulations, can effectively link simulation and experiment at the atomic scale. While atomic resolution scanning transmission electron microscopy (STEM) has made strides in filling this gap \cite{Kumar2021-gu}, conventional methods such as annular dark-field (ADF) and differential phase contrast (DPC) imaging project the three-dimensional (3D) crystal to a two-dimensional (2D) image, averaging information along the electron beam direction. The development of multislice electron ptychography presents an opportunity to overcome this limitation by recovering depth information. In this technique, diffraction patterns are collected for each probe scan position (Figure~\ref{fig:1}a). These patterns are then used to reconstruct the potential for each slice of the specimen along the beam direction (Figure~\ref{fig:1}b), providing 3D volumetric characterization of structure and chemistry \cite{Chen2021-rc, Harikrishnan2023-ly,Dong2024-vk}.

Here, the prototypical relaxor (1-$x$)\ce{PbMg_${1/3}$Nb_${2/3}$O_${3}$}–($x$)\ce{PbTiO_${3}$} (PMN-$x$PT, $x=0.32$) with the perovskite $AB$O$_{3}$ structure ($A$ = Pb$^{2+}$, \ce{B} = \ce{Mg$^{2+}$}, \ce{Nb$^{5+}$}, \ce{Ti$^{4+}$}) is examined using multislice electron ptychography. Direct 3D comparisons between experiment and MD simulations reveal the strong dependence of local polar textures on \ce{B} sub-lattice cations, influenced by their ionic radii and valence states. The overall disordered arrangement of these cations allows for local charge imbalance, disrupting dipole ordering and underpinning relaxor structure. 
Further examination of films under different compressive strain states shows that the influence of B sub-lattice chemistry remains significant along the constrained in-plane axes, while decoupling along the out-of-plane direction due to strain-induced polar displacement. The direct 3D comparison here bridges the gap between scattering experiments and MD simulations by providing an atomistic picture of the interplay between chemical inhomogeneities and strain in the prototypical relaxor-ferroelectric material system. By directly linking the structure determined from experiments to that of theoretical simulations, the complexities of nanoscale order and disorder are unraveled, demonstrating abundant opportunities for benchmarking and informing theoretical studies to address other complex alloys.

\section*{Polar Displacements in PMN-0.32PT}

Thin-film heterostructures of 55 nm \ce{0.68PbMg$_{1/3}$Nb$_{2/3}$O3}-\ce{0.32PbTiO3} (PMN-0.32PT)/25 nm \ce{Ba_{0.5}Sr_{0.5}RuO3} are grown on \ce{NdScO3}, \ce{Nd_{0.5}Sm_{0.5}ScO3}, and \ce{SmScO3} (110) substrates using pulsed-laser deposition, corresponding to epitaxy-imposed compressive strains of -0.50\%, -0.75\%, and -1.00\%, respectively (details in Ref.~\cite{Kim2022-xf} and SI).

\begin{figure}[!h]
\centering
  \includegraphics{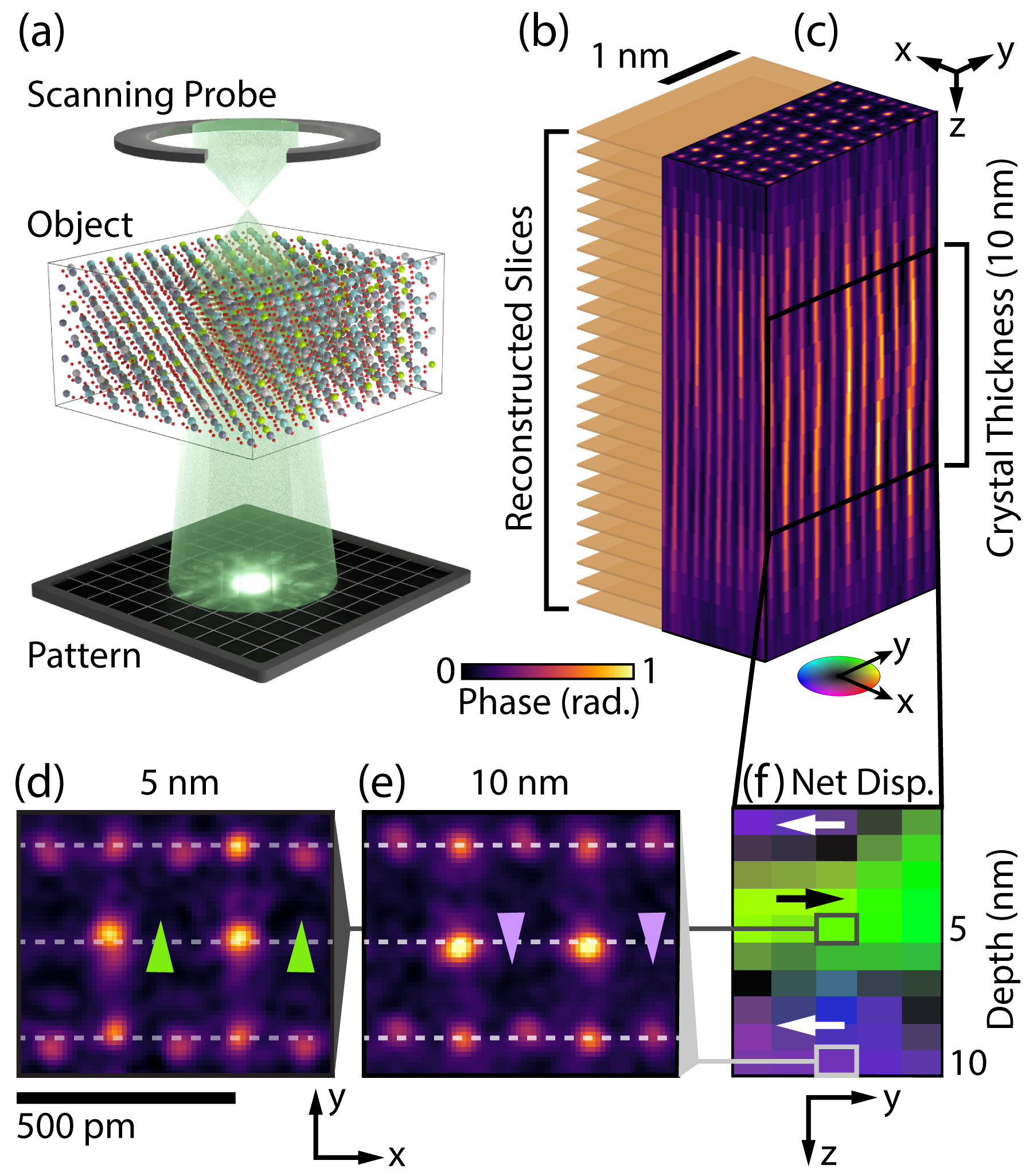}
  \caption{Schematic of data collection and analysis of volumetric polar displacements. (a) Overview of ptychography experiment and (b) slice-wise reconstruction (along z, zone axis of \hkl[110]) to recover the (c) projected potentials of the specimen for -1.00\% strain state PMN-0.32PT film. The z scale is compressed for visual simplicity. Two cropped slices are extracted from crystal thicknesses of (d) 5 nm and (e) 10 nm along the electron beam direction. Opposing polar displacements of the lead atom columns are marked. (f) Depth section along the plane in (c) showing three domains with changing net polar displacement. }
  \label{fig:1}
\end{figure}

The depth-resolved layers of the atomic structure are reconstructed using multislice electron ptychography for the -1.00\% compressed film ( Figure~\ref{fig:1}c). Each of the 23 slices represents the projected atomic potential within a 1 nm-thick region of the specimen along the electron beam direction. Because the surface layers exhibit damage from sample preparation, they are excluded from further analysis. Ten internal slices are then used for comparison to theory (SI-Figure S2).

Two representative slices from depths of 5 nm and 10 nm in the stack (Figure\ref{fig:1}d-e) show distinct positions of lead/oxygen, oxygen, and B sub-lattice atom columns, with varying contrast on the \ce{B} sub-lattice reflecting mixed occupancy by magnesium, niobium, and titanium. The extracted polar displacements (details in SI) reveal opposite shifts of 20 pm between the two depths, as schematically illustrated by the arrows. Note that, as the dwell time of a single electron probe position in the data acquisition is 1 ms, the ptychography reconstructions reflect the static, or average, structure. Repeating this analysis for all slices along the \hkl(1-10) cross sections provides direct measurements of the polar structure along the sample thickness direction (Figure~\ref{fig:1}f), revealing three domains in depth that would otherwise be overlooked with conventional projective imaging techniques \cite{Harikrishnan2023-ly}.

\begin{figure}[!h]
\centering
  \includegraphics[width=6.5in]{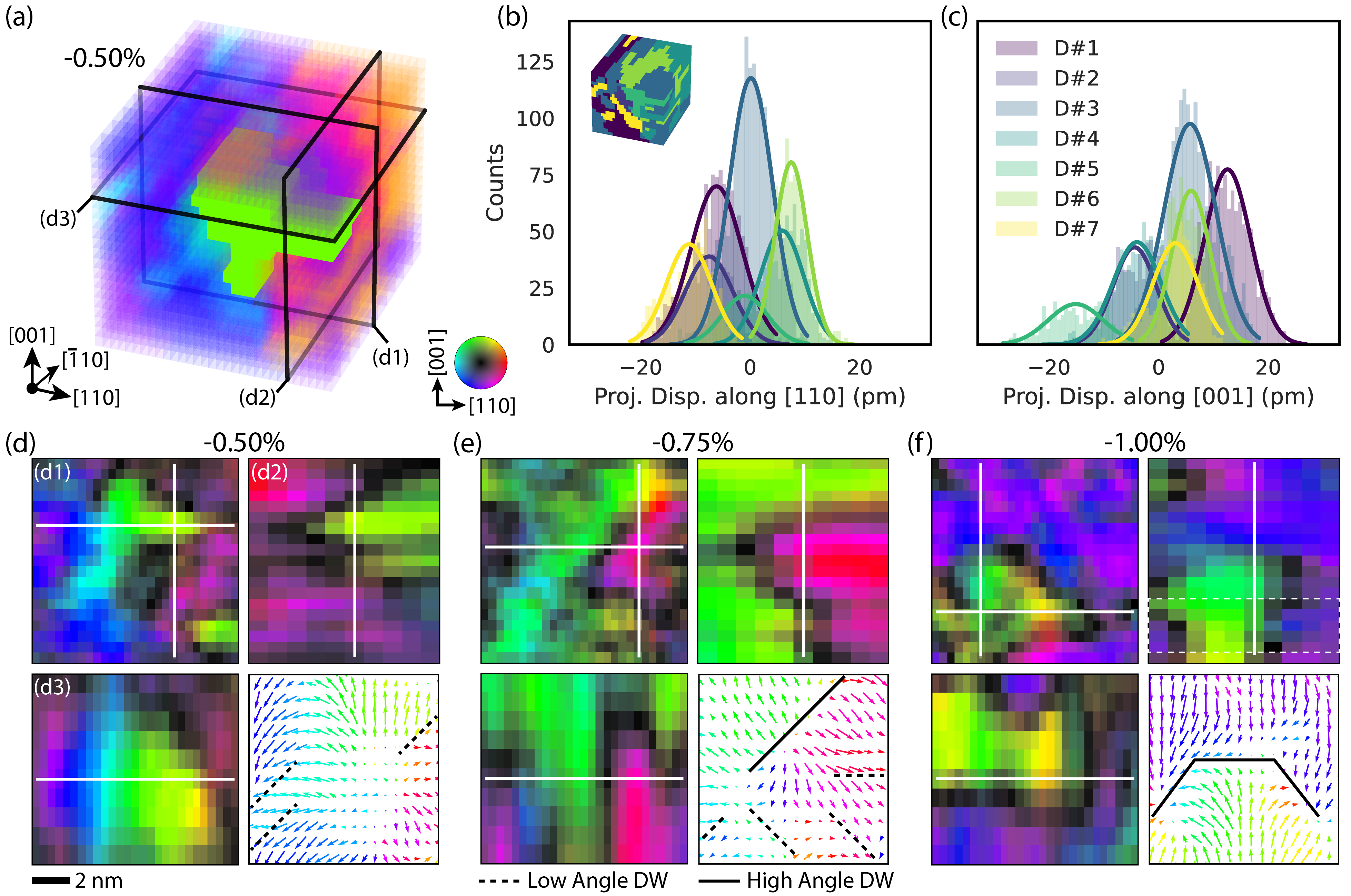}
  \caption{Polar domain structures of different strain states measured from reconstructions. (a) 3D view of the \hkl[110] and \hkl[001] polar displacements for a reconstructed -0.50\% ptychography stack. One example domain is shown with the solid color. The direction of polar displacements is indicated by the color wheel. (b) Projected \hkl[110] and (c) \hkl[001] displacement components from the dataset in (a), with domains clustered by similarity. (d) -0.50\%, (e) -0.75\%, and (f) -1.00\% slice and cross-section views from representative datasets showing the magnitude (saturation) and direction (color) of the \hkl[110] and \hkl[001] polar displacements, using the same color wheel as in (a). Sub-regions are shown by the respective quiver plots to better illustrate the flowing polar textures, with low-angle and high-angle domain walls indicated by dashed and solid lines, respectively. The dataset in (f) corresponds to that shown in Figure~\ref{fig:1}, with the marked cross-section matching Figure~\ref{fig:1}g. The saturation range of the color wheel represents a displacement magnitude interval of [0, 15] pm. }
  \label{fig:2}
\end{figure}

Using agglomerative clustering, depth-resolved polar displacement vectors for each lead-centered unit cell are grouped into domains (Figure \ref{fig:2}a, details in SI) for a -0.50\% strained dataset. A nanoscale multi-domain structure is observed, consisting of PNDs with gradually varying polar displacements on the order of 2-5 nm in size along all three dimensions (discussion of resolution in SI) \cite{Takenaka2013-df}. Within each PND, the broad distributions of the in-plane (IP, \hkl[110]) and out-of-plane (OOP, \hkl[001]) displacement components (Figure \ref{fig:2}b-c) suggest weak dipolar correlations within each domain. In addition, the near zero average of the overall displacement distribution reflects the relaxor-like random polarization distribution in the various domains (colors in Figure~\ref{fig:2}a) \cite{Takenaka2013-df}. From the section views and quiver plot visualization (Figure \ref{fig:2}d), the gradual variation of polarization is characterized by a high density of diffuse and low-angle domain walls, resembling a polar ``slush'' \cite{Takenaka2013-df, Takenaka2017-sj,Otonicar2020-at}. 

On top of the polar slush, the -0.50\% compressive strain condition induces a collective increase in OOP polarization, indicated by the slightly skewed distribution of the  displacement component (Figure \ref{fig:2}c vs b, SI-Figure S5). The increasing OOP displacement becomes more pronounced with higher strain values (-0.75\% and -1.00\%) (SI-Figure S5) \cite{Nagarajan2000-ki, Miao2016-qq, Belhadi2021-zc}.  Consequently, the domains grow and become more ferroelectric-like due to stronger polar correlations, accompanied by a transition from low-angle ($<$90\degree{}) to high-angle ($>$90\degree{}) domain walls as observed in the cross-section views (Figure~\ref{fig:2}e-f). Meanwhile, the epitaxial constraint has a marginal impact on the IP relaxor structure, as indicated by the similarly broad distribution of the IP displacement component (SI-Figure S5). In other words, the biaxial compressive strain preserves IP relaxor nature while enhancing OOP ferroelectric-like response \cite{Kim2019-pq, Kim2022-xf}.

The coexistence of relaxor- and enhanced ferroelectric-like ordering corroborates the evolution of X-ray diffuse scattering under increasing compressive strain \cite{Kim2019-pq, Kim2022-xf}. Here, precise depth-resolved measurements of polar displacement from ptychography further elucidate the strain-dependent modulation of polar order along different axes. The persistence of IP relaxor-like displacements even under -1.00\% strain suggests that certain intrinsic disorder in PMN-0.32PT remains unchanged even under external stimulus.

\section*{Polar Evolution in Simulations and Experiment}

In determining the origin of this disorder, the reported models for magnesium/niobium/titanium ordering point to chemical heterogeneity as a possible factor. Despite evidence from X-ray and neutron scattering and STEM HAADF imaging indicating increased chemical disorder beyond 25\% \ce{PbTiO3} in PMN-$x$PT \cite{Randall1990-fm, Krogstad2018-gl, Kumar2021-gu}, the relationship between chemical order and polar structure remains ambiguous in the absence of a 3D atomic-scale comparison. 
Here, MD simulations with varying biaxial compressive strain are conducted using two chemical order end cases, with one consisting of long-range rocksalt-ordered niobium (``random site'' model) and the other with disordered magnesium/niobium/titanium (details in SI)\cite{Akbas1997-yd,Hilton1990-eu,Krogstad2018-gl,Cabral2018-uv,Eremenko2019-wn,Kumar2021-gu}. Leveraging volumetric measurements of polarization from ptychography, these simulations are directly compared with experimental data to determine the impact of chemical heterogeneity on polar ordering, and to examine the types of ordering in PMN-0.32PT. Specifically, polar angle pair correlations as a function of distance (details in SI) are analyzed \cite{Takenaka2017-sj, Kim2019-pq, Kim2022-xf}, providing insights into polar domain size and structure.

\begin{figure}[!h]
\centering
  \includegraphics{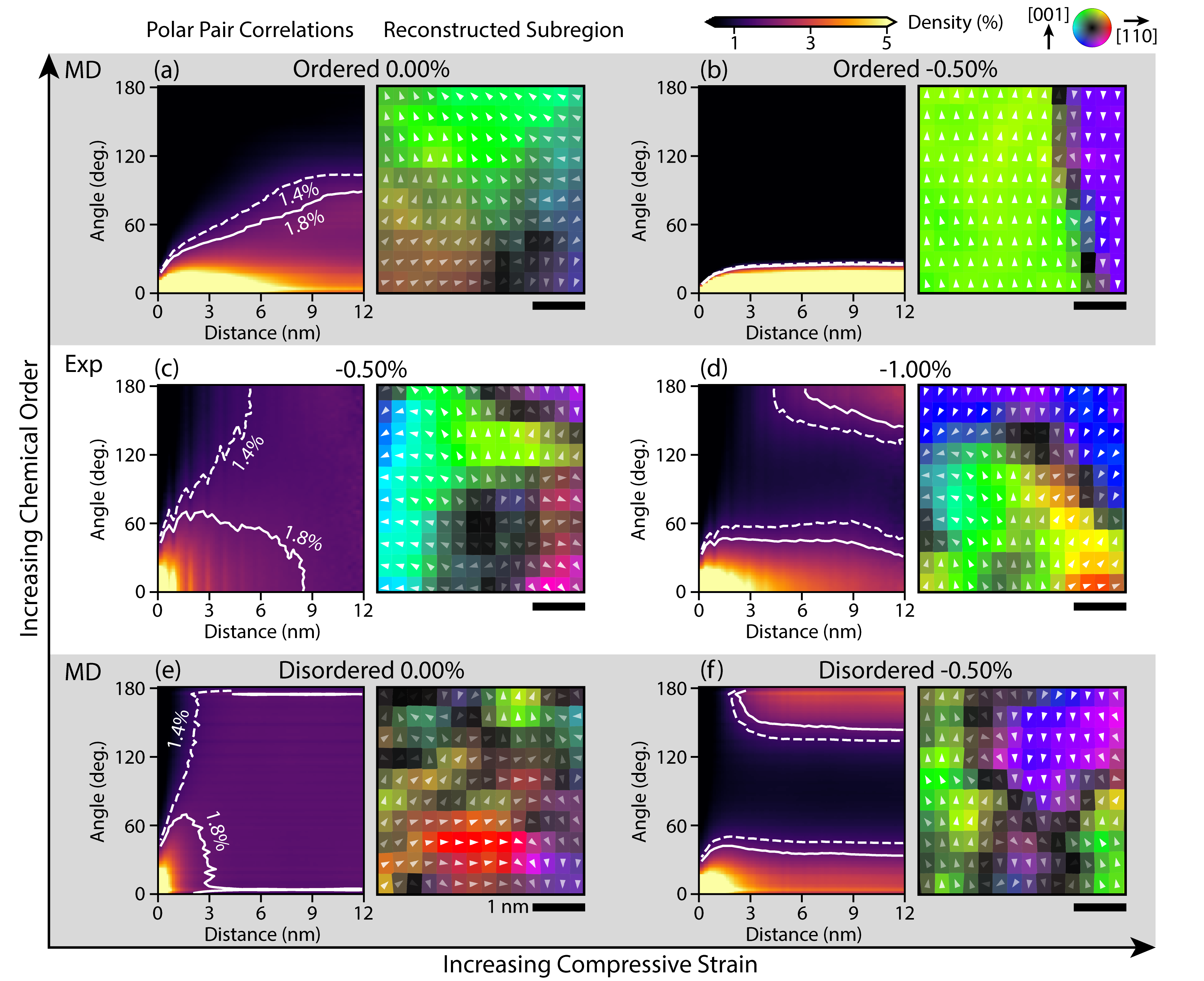}
  \caption{Polar angle pair correlation as a function of distance is calculated for MD structures with different configurations: (a) strain-free with charge-balanced \ce{Nb} order, (b) -0.50\% strain with charge-balanced \ce{Nb} order, (f) strain-free with random \ce{B} sub-lattice occupancy (disorder), and (g) -0.50\% strain with random B sub-lattice occupancy (disorder). For comparison, representative polar displacement maps are extracted from multislice simulations and ptychography reconstructions and displayed on the side. The correlation function is also calculated for experimental data on films under (c) -0.50\%, (d) -0.75\%, and (e) -1.00\% strain. The saturation range of the color wheel represents a displacement magnitude interval of [0, 15] pm. }
  \label{fig:3}
\end{figure}

In \textit{``random site''} MD structures with rocksalt-ordered niobium (Figure~\ref{fig:3}a), the map reveals strongly correlated polar dipoles (angle variation $<$30\degree{}) up to $\approx$6 nm. Beyond this range, the density gradually diminishes, indicating a distribution of PNDs with gradually varying polarization, typical of relaxor structure. This is further illustrated by the accompanying polar displacement map. The gradual polarization variation arises from the charge imbalance due to the random placement of magnesium and titanium atoms on the alternate rocksalt site not occupied by niobium. In the \textit{chemically disordered} MD structures (Figure~\ref{fig:3}e), however, local charge imbalance is further increased, favoring small uncorrelated PNDs and shortening the polar correlation length to $\leq$ 2 nm, as indicated by the plateau with near-uniform density beyond 2 nm \cite{Takenaka2017-sj, Zhang2024-uh}. This observation suggests that charge imbalance arising from the chemical disorder can aid in disrupting ferroelectric-like polar correlations, favoring the multi-domain relaxor-like structure. Application of compressive strain (Figure~\ref{fig:3}b vs a and f vs e), on the other hand, plays an opposite role in enhancing dipolar correlations by promoting ferroelectric-like behavior along \hkl[001], as illustrated with the accompanied local polar displacement maps (3D polar maps in SI-Figure S6).

Angle correlations of ptychography reconstructions from experiment (Figure~\ref{fig:3}c-d, respectively) closely resemble those of the \textit{chemically disordered} MD structures (Figure~\ref{fig:3}e-f). Beginning in the -0.50\% films, short-range polar angle correlations primarily cluster near 0\degree{} within $\approx$3nm, beyond which is a uniform density representing uncorrelated polarization. With strain increasing to -0.75\% (SI-Figure S12) and -1.00\% (Figure~\ref{fig:3}d), the near-random correlations segregate into two clusters at 0\degree{} and 180\degree{}, corresponding to domains with enhanced OOP polarization along either \hkl[001] or \hkl[00-1]. These clusters also become increasingly concentrated beyond 3 nm, consistent with the coarsening of PNDs to more ferroelectric-like ordered domains. At the same time, the number of high-angle domain walls between the predominantly \hkl[001]-aligned domains increases, as exemplified by the accompanying polar displacement map (Figure~\ref{fig:2}c-d).

In contrast, the strained and \textit{chemically ordered} MD structure (Figure~\ref{fig:3}b) presents a single concentrated maximum near 0\degree{} for dipole pairs up to 12 nm due to the enhanced polar correlations from the joint impact of strain and rocksalt ordering of niobium (SI-Figure S6). Furthermore, less pronounced charge imbalance in the chemical environment restricts the formation of high density charged domain walls, as exemplified by the head-to-tail polar displacement vectors (Figure~\ref{fig:3}b), rather than head-to-head polar displacements observed in experiments (Figure~\ref{fig:3}f). These charged regions, inherently higher in energy, can play an important role as pinning sites during domain coarsening and freezing at lower temperatures, thereby preventing relaxors from achieving complete ferroelectric order \cite{Takenaka2017-sj, Kim2022-xf}.

While the strain-induced domain evolution is captured by \textit{chemically disordered} MD structures, a systematically larger correlation length, or PND size, is observed in experiment ($>$3 nm in experiments vs. 2 nm in MD and also evidenced by the 1.8\% density contour, more discussion in SI). Such an enhancement of dipole alignment can be attributed to a greater degree of short-range ordered niobium and titanium in experiment compared to that found in the \textit{disordered} structure \cite{Kumar2021-gu, Krogstad2018-gl}, leading to stronger local polar correlations (Figure~\ref{fig:3}c-d vs. a-b). This is further supported by comparing the nearest-neighbor polar dipolar correlations for varying \ce{B} sub-lattice environments in the \textit{chemically disordered} MD structures (SI-Figure S13) \cite{Takenaka2017-sj, Kim2019-pq}. \ce{Pb}-\ce{O_12} dipoles surrounded with rocksalt-ordered niobium and titanium enhance local alignment compared to chemically disordered or magnesium-containing configurations, representing stronger short-range ferroelectric ordering \cite{Grinberg2009-mo, Juhas2004-jt, Grinberg2004-gx, Grinberg2012-cq}. The additional disorder on the \ce{B} sub-lattice determined from polar structure comparisons with ptychography bridges the overall heterogeneity seen in scattering experiments with atomic-scale simulations, providing a direct comparison and benchmark for models. Additionally, the mixture of long-range disorder with short-range order observed in experiments suggests that the \textit{local} influence of chemistry on polar structure must be further investigated.

\section*{Chemistry Mediated Local Polar Motifs}
\begin{figure}[!h]
\centering
  \includegraphics{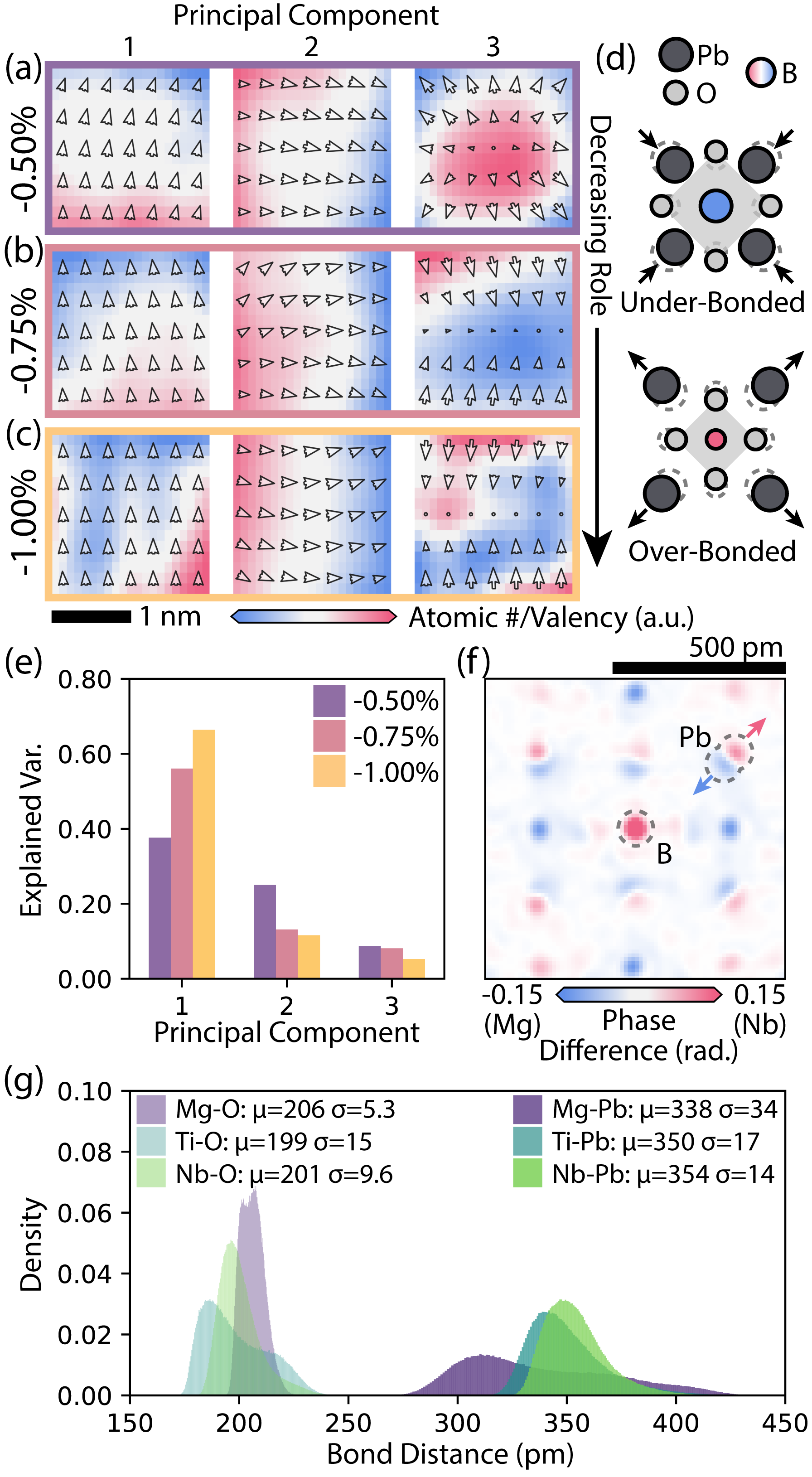}
  \caption{The principal components for polar textures within a 2 nm$^{3}$ window are extracted through PCA for grown thin films at different strain states. The first three components of (a) -0.50\%, (b) -0.75\%  and (c) -1.00\% films are visually represented with arrows overlaid on the chemical environment (details in SI). Note that only the positive weight (direction) of each PC is shown for simplicity. (d) Schematic illustrating the over/under-bonding effect on Pb displacements in (a)-(c). (e) The corresponding weight of each component in each sample is gauged by the explained variance. (f) Potential difference between the average of the highest 5\% versus lowest 5\% B sub-lattice motifs, with Pb displacements marked. (g) B-\ce{Pb} and B-\ce{O} bond lengths are extracted from strain-free MD structures with random B site occupancy.}
  \label{fig:4}
\end{figure}

Taking advantage of the depth-resolved atomic number sensitivity offered by ptychography, the influence of \ce{B} sub-lattice occupancy on local polar structure is further examined. First, the static, local polar environments are extracted from the ptychography reconstructions using overlapping $2\times 2\times 2$ nm$^3$ ``windows''. Next, principal component analysis (PCA) is performed to determine the principal patterns or motifs that best represent these local polar displacement windows, with their explained variances reflecting their weights or proportions in the films. 
For regions with minimal fluctuations (\textit{e.g.} ferroelectric-like domains), much of the local polar structure can be expressed as a linear combination of orthogonal positive or negative components along \hkl[001] and \hkl[110] (PC\#1 and \#2, Figure \ref{fig:4}a-c respectively). The higher-order components, on the other hand, represent local polarization fluctuations and domain walls (SI-Figure S9). Compared to strain-free PMN-0.32PT with a global rhombohedral structure \cite{Krogstad2018-gl, Kim2022-xf}, in which \hkl<111> displacements can be represented as equal proportions of PC\#1 (\hkl[001]) and \#2 (\hkl[110]), the increasing weight of PC\#1 in strained films (Figure \ref{fig:4}e) reflects the enhanced OOP polarization. In turn, the weights of higher-order components decrease at higher strain states as the PNDs coarsen to larger ferroelectric-like domains (details in SI).

To examine the influence of the chemical environment on these local polar motifs, the average integrated potential ($I_{P}$) of the \ce{B} sub-lattice atom columns is extracted for each PC (details in SI). This potential is proportional to atomic number ($Z$)\cite{Chen2021-rc} and, in this particular case, the valence state of the \ce{B} sub-lattice species, from \ce{Mg^{2+}} ($Z$ =12) to \ce{Ti^{4+}} ($Z$ =22) to \ce{Nb^{5+}} (Z = 41). The observed relationship between polar motifs (polar displacements represented with arrows) and chemical environment ($I_P$, integrated potential represented with background colors, Figure \ref{fig:4}a-c) suggests that atomic regions with lower valence state (\textit{e.g.}, \ce{Mg^{2+}}-rich regions) act as ``sinks'' for polarization, to which surrounding polar displacements converge. Conversely, regions with higher valence state (\textit{e.g.,} \ce{Nb^{5+}}-rich regions) serve as ``sources'' from which polar displacements originate. This effect, rooted in charge imbalance and underbonding/overbonding of oxygen anions \cite{Juhas2004-jt, Grinberg2007-xx, Takenaka2013-df}, is schematically illustrated (Figure~\ref{fig:4}d) and directly visible from ptychography reconstructions (Figure~\ref{fig:4}f).

Comparing the PCA-derived motifs from experiment with those from MD structures shows a consistent dependency of polarization on the local chemical environment (details in SI). This observation can be rationalized in two parts. First, as the canonical charge increases (more \ce{Nb^{5+}}), a stronger Coulombic force repels \ce{Pb^{2+}}, positioning the positive charge center further from the \ce{B} sub-lattice. This is seen in the average \ce{Pb}-\ce{B} bond lengths extracted from the MD structures, which are the longest when \ce{B} = \ce{Nb^{5+}} (Figure~\ref{fig:4}g). Second, with decreasing ionic radii of the \ce{B} sub-lattice species (\ce{Mg^{2+}} to \ce{Nb^{5+}} to \ce{Ti^{4+}}), the \ce{B}-\ce{O6} octahedra bond lengths decrease (Figure~\ref{fig:4}g) \cite{Grinberg2004-gx, Grinberg2002-lm}. In other words, smaller cations, like \ce{Nb^{5+}}, are more tightly bound to their neighboring anions and are therefore positioned closer to the negative charge center. As a consequence, the offset between anion and cation centers leads to a net polarization pointing away from \ce{Nb^{5+}} and towards \ce{Mg^{2+}}. 

Despite intuitive interpretation, in chemically disordered relaxors, the strong correlation between polarization and chemistry at the atomic scale can be underestimated or even overlooked without a precise probe of local chemistry \cite{Kumar2021-gu}. The observations here, enabled by depth-resolved volumetric characterization, underscore the influence of charge state and ionic radius on bond-length distributions and ultimately polarization in relaxors, while highlighting the potential of designing relaxor-ferroelectric response through manipulation of \ce{B} sub-lattice chemistry.

\section*{Joint Impact of Strain and Local Chemistry}
Both strain and chemical environment are shown to significantly influence the evolution of polarization in PMN-0.32PT. Locally, the relaxor-like polar structure is promoted by charge imbalance resulting from chemical heterogeneity, with the strength of polar correlations dictated by the randomness of the \ce{B} sub-lattice. The incorporation of more \ce{PbTiO3} disturbs the rocksalt-ordered niobium in PMN-0.25PT (compared to PMN-0.32PT here) and leads to smaller and less correlated PNDs by introducing stronger local charge imbalance \cite{Takenaka2017-sj}. The varied bond characteristics of disordered magnesium/niobium/titanium introduce both paraelectric- and ferroelectric-like distortions arising from bonding and interactions with \ce{O^{2-}} and \ce{Pb^{2+}} \cite{Grinberg2004-gx, Matsuura2006-aa,Kamba2007-uj, Grinberg2009-mo,Grinberg2012-cq}. Together, these structural factors contribute to dielectric relaxation and the foundation of relaxor behavior \cite{Sepliarsky2011-kd, Takenaka2013-df, Takenaka2017-sj, Bokov2011-jv}. At the mesoscale, clusters with specific \ce{B} sub-lattice occupancy, such as rocksalt-ordered niobium, exhibit greater polar correlations (SI-Figure S13) or ferroelectric-like order \cite{Juhas2004-jt,Pasciak2012-bm, Grinberg2009-mo}. Conversely, regions with more weakly-coupled or more paraelectric-like fluctuations serve to break long-range order, resulting in the PND ``slush'' \cite{Grinberg2007-xx, Takenaka2013-df, Takenaka2017-sj}. While the length scale of PNDs is affected by chemical inhomogeneity, it is not directly coupled to unit-cell-level chemical oscillations but instead is influenced by variations in ionic radii and valency of the overall chemical environment.

Finally, across the global (film) scale, the application of strain mirrors the effect of electric field by inducing long-range polar correlations and enhancing out-of-plane ferroelectric response \cite{Kim2019-pq,Kim2022-xf}. This transformation develops on top of the existing polar slush without negating chemical contributions \cite{Xu2006-jo}. Looking again at disordered MD structures, higher compressive strain results in a larger \hkl[001] polarization component (SI-Figure S8), simplifying the primary local polar motifs and mirroring observations from experiment (Figure~\ref{fig:4}e). Within the constrained dimension, however, the \hkl[100] and \hkl[010] polarization components remain balanced at all compressive strain states (SI-Figure S7 and S10), preserving the chemical contribution and resembling pressure-induced relaxor response \cite{Samara1996-yj, Cohen1992-dn, Tinte2006-pp}. This finding demonstrates the ability of strain to selectively induce polarization along \hkl[001], with potential applications through mechanical and chemical constraints \cite{Gao2021-fl}. 

\section*{Conclusion}

Through direct, volumetric comparisons between MD simulations and experiment using multislice electron ptychography, the connections between 3D polar structure, chemical environment, and strain in a relaxor ferroelectric have been determined. The relaxor-like polar structure, characterized by continuous changes in polar correlations, is shown to exhibit strong coupling with local charge imbalance due to chemical heterogeneities. Analysis of 3D polar motifs and local chemical environments from experiment and MD simulations consistently indicates that lower valence state species (\textit{e.g.}, \ce{Mg^{2+}}) act as sinks for polarization, while higher valence state species (\textit{e.g.,} \ce{Nb^{5+}}) act as sources. The random occupancy of these ions on the \ce{B} sub-lattice collectively impacts polar structure, underpinning the relaxor-like structure. Biaxial compressive strain, in turn, enhances out-of-plane polarization without negating the influence of chemistry, resulting in the coexistence of ferroelectric- and relaxor-like order. 

As a whole, the ability to probe the full complexity of nanoscale order and disorder using multislice electron ptychography enables direct 3D comparisons with structures determined from simulation. By bridging experiment and theory, these results both demonstrate a benchmark for modeling relaxor structure and behavior and also offer insights into the correlative study and hierarchical control of polar structure in PMN-$x$PT through chemical order and strain. This approach presents significant opportunities for tackling other complex materials through complementary structural characterization and simulation at the atomic scale. 

\section*{Acknowledgements}
This research was sponsored by the Army Research Laboratory and was accomplished under Cooperative Agreement Number W911NF-24-2-0100. The views and conclusions contained in this document and those of the authors should not be interpreted as representing the official policies, either expressed or implied, of the Army Research Laboratory or the U.S. Government. The U.S. Government is authorized to reproduce and distribute reprints for Government purposes, notwithstanding any copyright notation herein.  Y.Q., J. Z., and A.M.R. acknowledge support for the theoretical and computational research by the Office of Naval Research, under grant N00014-24-1-2500.  Computational support was provided by the High-Performance Computing Modernization Program of the U. S. Department of Defense. J.K. acknowledges the support of the Army Research Office under the ETHOS MURI via cooperative agreement W911NF-21-2-0162. B.R.D. acknowledges the support of the National Science Foundation Graduate Research Fellowship under Grant No. 2141064. This work made use of the MIT.nano Characterization Facilities.

\section*{Author Information} 
\subsection*{Contributions}
M.Z. and M.X. conducted electron microscopy experiments. M.Z., M.X., and C.G. conducted data analysis and image simulations. Y.Q. and J.Z. performed MD calculations. J.M.L., A.M.R., and L.W.M. designed and supervised this research. All authors co-wrote and edited the manuscript.

\section*{Ethics declarations}
\subsection*{Competing Interests}
The authors declare no competing interests.

\section*{Supporting Information} 
Supporting Information is available from the author.

\section*{Data Availability} 
The 4D-STEM dataset, hyperparameters used for reconstruction, and the final output can be accessed at \url{Upload to Zenodo}. Other data and codes for image analysis are available from the corresponding author by reasonable request.

\medskip
\bibliographystyle{Science}
\bibliography{paperpile.bib}

\newpage

\end{document}